\documentclass[journal=amlccd,manuscript=letter]{achemso}

\usepackage{chemformula} % Formula subscripts using \ch{}
\usepackage[T1]{fontenc} % Use modern font encodings
\usepackage{graphicx}
\usepackage{color}
\usepackage{bm}
\usepackage{physics}
\usepackage{amsmath, amssymb}

\graphicspath{{./}{./figs/}{../figs/}}

\newcommand{\mean}[1]{\left\langle #1 \right\rangle}

\author{Shota Goto}
\affiliation{Division of Chemical Engineering, Department of Materials Engineering Science, Graduate School of Engineering Science, Osaka University, Toyonaka, Osaka 560-8531, Japan}

\author{Takenobu Nakamura}
\affiliation{National Institute of Advanced Industrial Science and Technology (AIST), 1-1-1 Umezono, Tsukuba, Ibaraki 305-8568, Japan}

\author{Davide Michieletto}
\affiliation{School of Physics and Astronomy, University of
Edinburgh, Peter Guthrie Tait Road, Edinburgh, EH9 3FD, UK}
\alsoaffiliation{MRC Human Genetics Unit, Institute of Genetics and Cancer, University of Edinburgh, Edinburgh, EH4 2XU, UK}

\author{Kang Kim}
\email{kk@cheng.es.osaka-u.ac.jp}
\affiliation{Division of Chemical Engineering, Department of Materials Engineering Science, Graduate School of Engineering Science, Osaka University, Toyonaka, Osaka 560-8531, Japan}

\author{Nobuyuki Matubayasi}
\email{nobuyuki@cheng.es.osaka-u.ac.jp}
\affiliation{Division of Chemical Engineering, Department of Materials Engineering Science, Graduate School of Engineering Science, Osaka University, Toyonaka, Osaka 560-8531, Japan}

\title[Persistent Homology of Topological Glasses]
{Persistent Homology Reveals the Role of Stiffness in Forming Topological Glasses in Dense Solutions of Ring Polymers}

\begin{document}

\begin{tocentry}
\centering
\includegraphics[scale=0.8]{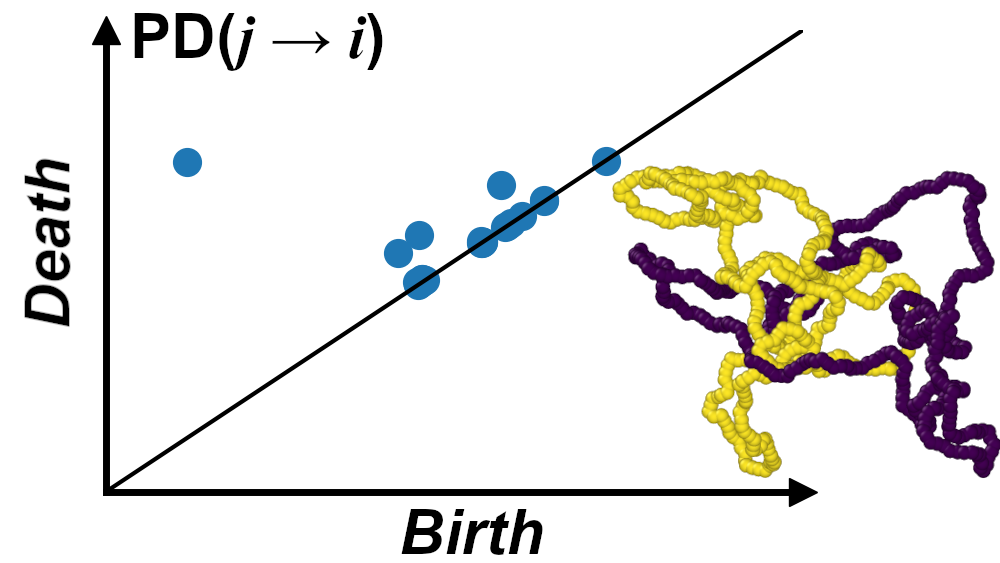}
\end{tocentry}

\begin{abstract}
Ring polymers are characterized by topology-specific entanglements called threadings. 
 In the limit of large rings, it is conjectured that a ``topological glass''
should emerge due to the proliferation of threadings.
 In this study, we used persistent homology to quantify threading structures of ring polymers 
with different chain stiffness
and elucidate mechanisms behind topological glasses. 
Using coordination data from coarse-grained
 molecular dynamics simulations, 
we analyzed the topology of the union of virtual spheres centered on each monomer or center of mass.
As the radius of each sphere increases, the corresponding points
 connect, giving rise to topological entities such as edges, loops, and
 facets.
We then analyzed how the number of loops per ring chain
 and penetrated loops varies with sphere radius, focusing on the effects
 of chain stiffness and density.
The results reveal that loops are larger in stiff
 ring chains, whereas flexible ring chains do not generate
 sufficiently large loops to establish a threading structure.
The stiffness of ring polymer plays a significant role in
 the formation of topological glasses in ring polymers.
\end{abstract}

Ring polymers exhibit distinctive properties compared to
their linear counterparts~\cite{cates1986Conjectures,
muller1996Topological, grosberg1988Role, sakaue2011Ring}. 
Despite extensive research, 
a thorough
understanding of topological constraints in ring
polymer melts remains a significant challenge in polymer
physics~\cite{suzuki2009Dimensiona, tsolou2010Melt, halverson2011Molecular, halverson2011Moleculara,
halverson2012Rheology, bernabei2013Fluids, pasquino2013Viscosity, bras2014Compact, slimani2014Clustera,
ge2016SelfSimilar, tsalikis2017Microscopic, iwamoto2018Conformationsa, mei2020Microscopic,
jeong2020Intrinsic, kruteva2020SelfSimilar,  
mei2021Theorya, goto2021Effects, smrek2020Active, stano2022Thread,
chubak2022Active, roy2022Effect, cai2022Conformation, tu2023Unexpected, kruteva2023Topology, mei2024Unified}. 
One key feature thought to define topological constraints in ring polymers is 
the interpenetrating structure known as ``threading''.
Threading occurs when one ring
polymer penetrates the loop of another ring polymer, with the
penetrating ring classified as active and the penetrated ring as
passive, 
illustrating the asymmetric and 
hierarchical nature of the threading network.
For sufficiently long rings, 
this threading network can evolve, eventually 
leading to the formation of ``topological glasses,'' where the 
relaxation time
is expected to increase drastically with respect to the extent of
threading~\cite{lo2013Topological, michieletto2016Topologically,
michieletto2017Ring, sakaue2018Topological}.

Analyzing threading and clarifying its relationship
with glass-like properties is crucial.
While several approaches for quantifying threading have been proposed, including
methods based on minimal surface~\cite{smrek2016Minimal, smrek2019Threading} and geometric
analysis~\cite{tsalikis2014Threading, tsalikis2016Analysis}, Landuzzi et
al. introduced a method for quantifying the
threading of ring polymers using persistent homology (PH)~\cite{landuzzi2020Persistence}.
PH is a mathematical tool that characterize topological features
such as ``loops'' from point cloud~\cite{edelsbrunner2000Topological, edelsbrunner2010Computational,
hiraoka2016Hierarchical}.
Specifically, Landuzzi et al. investigated threading structures using PH from 
MD simulations with the Kremer--Grest (KG) model~\cite{kremer1990Dynamics} for
ring polymers.
Of particular interest was 
the chain length $N$ dependence of ring polymers up to $N=2048$ at a monomer number
density of 0.1, incorporating a bending potential
$U_\mathrm{bend}(\theta) = \varepsilon_\theta(1+\cos\theta)$, where
$\theta$ represents the angle formed by consecutive bonds and 
$\varepsilon_\theta = 5$ (see Eq.~\eqref{eq:bending} for details).
This bending potential effectively models the polymers as worm-like
chains, analogous to 
the Kratky--Porod model~\cite{kratky1949Rontgenuntersuchung}.

We recently performed MD simulations using the KG
with two types of ring polymers: semi-flexible ($\varepsilon_\theta =
1.5$) and stiff ($\varepsilon_\theta = 5$) rings with a fixed chain length
$N=400$ to
investigate the influence of chain stiffness on their dynamic
properties~\cite{goto2023Unraveling}.
The rearrangement dynamics of the center of mass (COM) were analyzed,
with a focus on dynamic heterogeneity to clarify glassy behavior.
Our results demonstrated that stiff ring polymers exhibit
pronounced glassy behavior accompanied by dynamic heterogeneity, whereas
semi-flexible ring polymers display homogeneous dynamics characterized
by a Gaussian distribution of COM displacement.
This distinction suggests
that the dynamic properties of ring polymers are fundamentally
influenced by the chain stiffness, emphasizing the need to examine
threading structures across varying degrees of chain stiffness.

The purpose of this study is to elucidate the influence of the chain
stiffness and monomer number density of ring polymers on their threading structures. 
We first analyze the connectivity of COM using PH.
Subsequently, 
we characterize the 
active and passive threading structures between pairs of ring chains through PH.
Through these analyses, we clarify the topological
characteristics of ring polymers and their relationship to glassy
behavior, informed by insights gained from the rearrangement dynamics of
COM.

%%%%%%%% Methods %%%%%%%%
{\noindent \sffamily \bfseries Model and Methodology\\}
We employed MD simulations for ring polymer dense solutions utilizing the KG model.
Each ring polymer is represented by $N$ monomer beads, each with mass
$m$ and diameter $\sigma$.
The system comprises $M$ ring chains contained within a
three-dimensional cubic box with volume of $V$, with periodic boundary conditions.
The monomer beads interact through
three types of inter-particle potentials:
the Lennard-Jones (LJ) potential governs the interaction between 
all pairs of monomer beads and is defined as 
\begin{equation}
  U_{\mathrm{LJ}}(r) = 4 \varepsilon_{\mathrm{LJ}}
  \qty[ \qty(\frac{\sigma}{r})^{12} - \qty(\frac{\sigma}{r})^6 ]
  + C, 
  \label{eq:LJ}
\end{equation}
where $r$ is the distance between two beads, 
$\varepsilon_\mathrm{LJ}$ is the depth of the potential well, and $C$ 
is a constant that shifts the potential at the cut-off distance of $r_\mathrm{c} = 2^{1/6}\ \sigma$.
Two adjacent monomer beads along the chain also interacted
via the finitely extensible nonlinear elastic (FENE) bond potential
\begin{equation}
  U_{\mathrm{LENE}}(r) = -\frac{1}{2} K R_0^2
  \ln \qty[1 - \qty(\frac{r}{R_0})^2]
  \label{eq:FENE},
\end{equation}
for $r < R_0$, where $K$ and $R_0$ represent the spring constant and the 
maximum length of the bond, respectively.
%Note that Eqs~\eqref{eq:LJ} and \eqref{eq:FENE} define
%the finitely extensible nonlinear elastic (FENE) bond potential of the
%KG model.
We used the values of $K=30 \varepsilon_\mathrm{LJ}/\sigma^2$ and $R_0 =1.5 \sigma$.
Lastly, 
the chain stiffness is controlled by incorporating a bending potential
\begin{equation}
  U_{\mathrm{bend}}(\theta) = \varepsilon_\theta 
  \qty[1 - \cos (\theta - \theta_0)],
  \label{eq:bending}
\end{equation}
where $\theta$ is the bending angle formed by three consecutive monomer beads
along the polymer chain.
In this study, the bending energy was set as 
$\varepsilon_\theta/ \varepsilon_{\mathrm{LJ}} = 0$, 1.5, 2, 3, 4, and 5, 
with an equilibrium angle of $\theta_0 = 180^\circ$.

All MD simulations 
%in this study 
were 
%conducted 
performed using the Large-scale Atomic/Molecular
Massively Parallel Simulator (LAMMPS)~\cite{plimpton1995Fast}.
Length, energy and time are represented in units of $\sigma$, $\varepsilon_\theta$ and 
$(m / \varepsilon_\mathrm{LJ})^{1/2}$, respectively.
Additionally, the temperature is expressed in units of $\varepsilon_\mathrm{LJ} / k_\mathrm{B}$, 
where $k_\mathrm{B}$ is Boltzmann constant.
We fixed the temperature $T$, chain length $N$, number of chains $M$ as $T=1.0$ and
$N=400$, and $M=100$, respectively.
Throughout the simulations,
temperature was controlled using 
the Nos\'{e}--Hoover thermostat, with a time step of $\Delta t = 0.01$.
The monomer number density $\rho\sigma^3$ ($= NM\sigma^3/V$) was varied as 0.1, 0.2, 0.3, 0.4,
and $0.5$ for each degree of chain stiffness.
%Note that the common value of $\rho=0.85$ was excluded because 
%nematic ordering was observed for stiff ring chains with
%$\varepsilon_\theta = 5$.
Henceforth, 
$\rho$ will be referred to as density.

\begin{figure}[t]
\centering
\includegraphics[width=\textwidth]{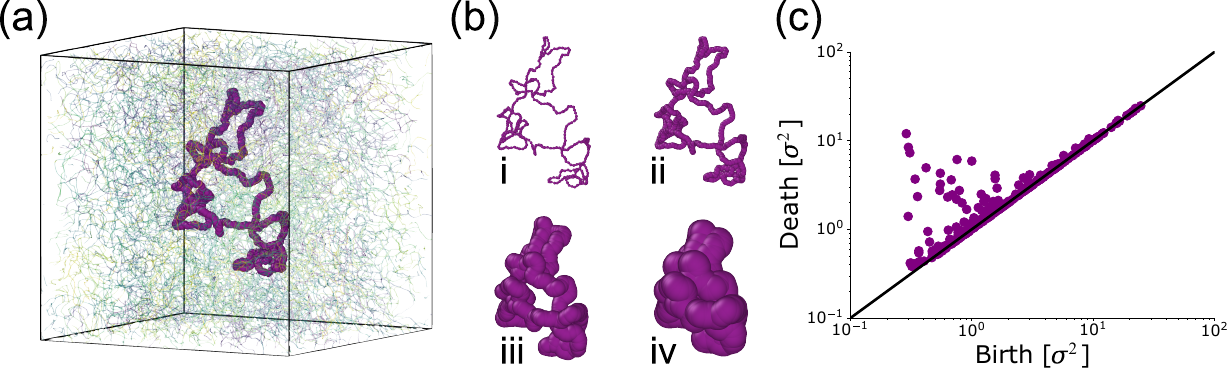}
\caption{
Persistent diagram of an ring polymer
    with chain length $N=400$ at density $\rho=0.1$.
    (a) Snapshot of entangled ring polymer chains.
    (b) Virtual spheres with radii $\sqrt{\alpha}$ assigned on each monomer bead of an arbitary 
    ring.
    (c) Persistent diagram of the ring.
}
\label{fig:ph_mockup}
\end{figure}

%\revisionBlue{Let us now provide a brief outline of PH.}
%We employed PH, a mathematical method used to \revision{analyze} point
%and generate Persistent Diagrams (PD), 
%to investigate the complex interpenetrating structures in ring polymer systems.
Here, we briefly outline PH:
A set of coordinates such as beads of chains or COM, 
denoted as
$\{\bm{r}\} = \{\bm{r}_1, \bm{r}_2, \ldots, \bm{r}_\kappa\}$, is used as input data,
where $\kappa$ is the number of coordinates.
At each coordinate, assign a virtual sphere with radius $\sqrt{\alpha}$,
where $\alpha$ is a parameter. 
Initially, when $\alpha=0$, all points are treated as disconnected
components. 
As $\alpha$ increases, the spheres begin to overlap, and connected
components form, creating edges and facets.
During this process,
%``birth'' and ``death'' of 
the topological features varies discontinuously with
respect to $\alpha$, \textit{i.e.}, the loops will appear and
disappear.
%The scales of birth and death are denoted as 
%``$b$''
%and 
%``$d$''
%, respectively, and are visualized in a two-dimensional diagram, known as PD. 
We record the radii for appearance and disappearance as $b$ (birth) and $d$ (death) respectively for each hole, 
and introduce persistence diagram (PD) as a collection $(b,~d)$ of all
holes. 
In this context, a zero-dimensional hole represents a connected
component, while
a one-dimensional hole represents a loop.
The instance of the PH procedure is illustrated in Fig.~\ref{fig:ph_mockup},
which shows the PD of a ring polymer in solution.

%This PD allows for the identification of topological structures that evolve with changes in $\alpha$
PD captures not only the topological features at a specific radius, \textit{i.e.}, a threshold of connection, but also how these features change as the threshold increases.
All analysis were performed using the HomCloud~\cite{obayashi2022Persistent}.

%%%%%%%%% Fig. 1 %%%%%%%%%%%%%%%%%%%%%%%%%%%%%%%%%%%%%%%%%%%%%%%%%%%%%%%%%%%%%%%%%%%%%%%%%
%%%%%%%%%%%%%%%%%%%%%%%%%%%%%%%%%%%%%%%%%%%%%%%%%%%%%%%%%%%%%%%%%%%%%%%%%%%%%%%%%%%%%%%%%%

%%%%%%%%% Fig. 2 %%%%%%%%%%%%%%%%%%%%%%%%%%%%%%%%%%%%%%%%%%%%%%%%%%%%%%%%%%%%%%%%%%%%%%%%%
\begin{figure}[t]
\centering
\includegraphics[width=\textwidth]{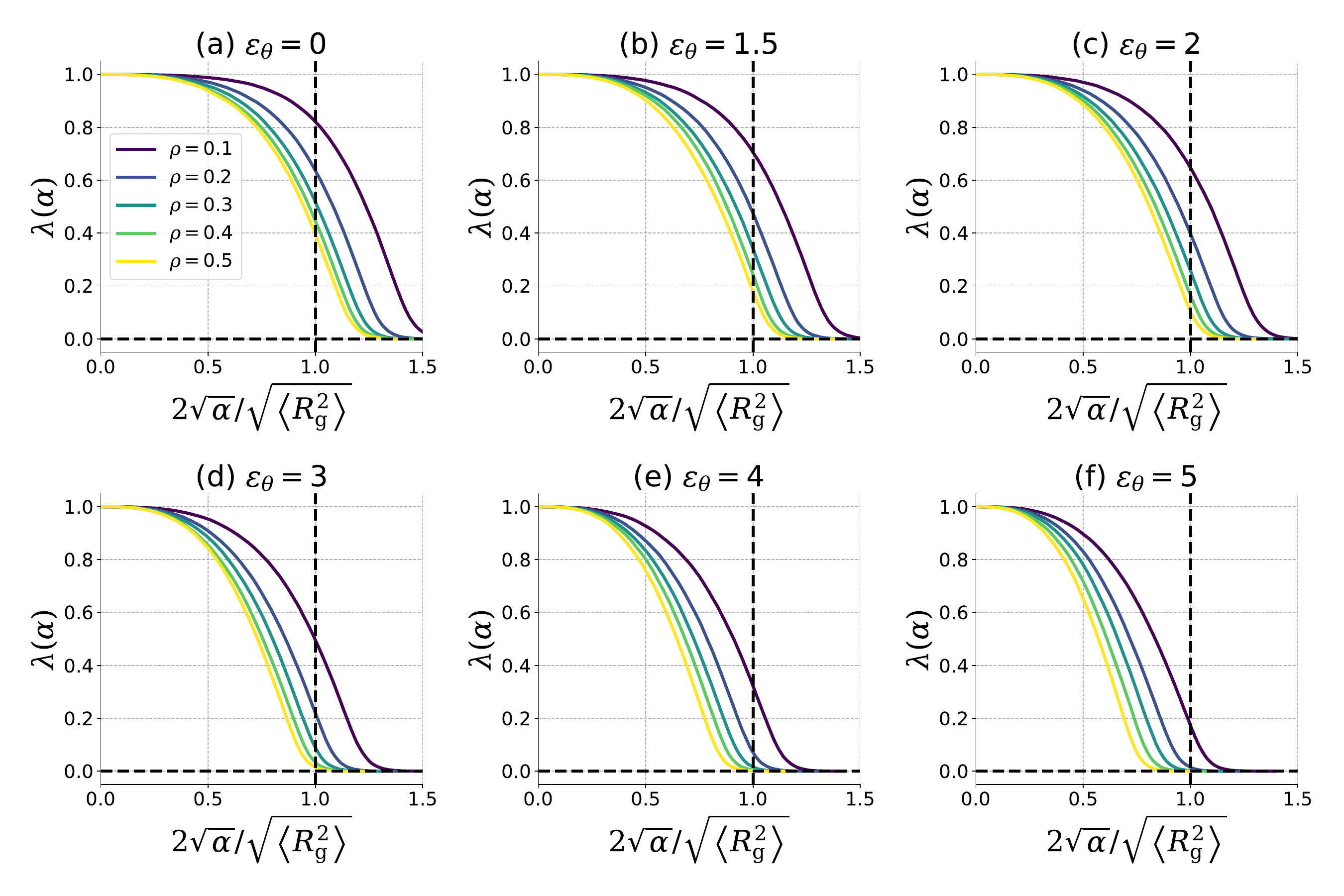}
\caption{Density $\rho$ dependence of 
$\lambda(\alpha)$ as a function of
 $2\sqrt{\alpha}/\sqrt{\mean{R_\mathrm{g}^2}}$ by varying the bending energy
 $\varepsilon_\theta = 0$ (a),  $\varepsilon_\theta = 1.5$ (b),
 $\varepsilon_\theta = 2$ (c),  $\varepsilon_\theta = 3$ (d), 
 $\varepsilon_\theta = 4$ (e),  and $\varepsilon_\theta = 5$ (f).
The horizontal and vertical dashed lines represent
$\lambda=0$ and $\alpha=\mean{R_\mathrm{g}^2}/2$, respectively.
}
\label{fig:ph_com}
\end{figure}
%%%%%%%%%%%%%%%%%%%%%%%%%%%%%%%%%%%%%%%%%%%%%%%%%%%%%%%%%%%%%%%%%%%%%%%%%%%%%%%%%%%%%%%%%%

{\noindent \sffamily \bfseries Connectivity of Centers of Mass in Ring Polymer Solutions\\}
%%%%%%%% Results %%%%%%%%
The first analysis aims to reveal the connectivity of the COM
coordinates of ring polymer chains using PH.
%PH quantifies the
%number of connected components of point coordinates with increasing the
%parameter $\alpha$.
%The number of connected components at a given $\alpha$ is denoted as
%$\beta_0(\alpha)$, which is referred to as 
%the zero-th Betti number.
The number of connected components at a given $\alpha$, denoted as $\beta_0(\alpha)$ and referred to as the zero-th Betti number, is calculated.
As the radius of the virtual sphere with a radius of $\sqrt{\alpha}$ expands, 
%overlapping spheres are treated as a single connected component.
spheres will connect each other and finally become one lump.
%Eventually, the number of connected components 
Thus, $\beta_0(\alpha)$ converges to unity as $\alpha$
approaches infinity.
We define the function representing the decrease in
$\beta_0(\alpha)$ as
\begin{equation}
\lambda(\alpha) = \mean{\frac{\beta_0(\alpha) - 1}{\beta_0(0)-1}},
\label{eq:lambda_alpha}
\end{equation}
where $\mean{\cdots}$ represents the statistical average over the snapshot
configurations generated by MD simulations; note that $\beta_0 (\alpha)$ can be 
determined for each snapshot.
Accordingly, 
this function takes the value 
$\lambda(\alpha=0)=1$ and $\lambda(\alpha\to \infty) = 0$.
The density $\rho$ dependence of $\lambda(\alpha)$ is plotted in
Fig.~\ref{fig:ph_com} by varying the bending energy $\varepsilon_\theta$.
In the plot, the horizontal axis is represented by 
$2\sqrt{\alpha}/\sqrt{\mean{R_\mathrm{g}^2}}$ with the 
mean square gyration of radius
$\mean{R_\mathrm{g}^2}$ of the ring chains.
Note that the COM distance between any pair of 
two ring polymers, $i$ and $j$, is related as
$r_{ij}=2\sqrt{\alpha}$ when the rings are in contact, since
the radius of the sphere in the PH analysis is 
$\sqrt{\alpha}$.

Figure~\ref{fig:ph_com} demonstrates that $\lambda(\alpha)$
decreases and converges to zero at a specific length scale $\alpha$. 
This behavior indicates percolation, where clusters are formed by virtually
connected COMs. 
The characteristic length scale is $\alpha = \mean{R_\mathrm{g}^2}/2$, where 
a virtual bond is considered to have formed if the distance $r_{ij}$ between
the COMs of ring polymer pair $(i, j)$ satisfies $r_{ij} \le
\mean{R_\mathrm{g}^2}$ (see horizontal lines in Fig.~\ref{fig:ph_com}).
For flexible ring chains with $\varepsilon_\theta = 0$, $\lambda$ takes finite values for $\alpha
\le \mean{R_\mathrm{g}^2}$ across all densities, indicating the presence of
numerous small clusters. 
In contrast, stiff ring chains with $\varepsilon_\theta = 5$ exhibit
$\lambda \approx 0$ at $\alpha =
\mean{R_\mathrm{g}^2}/2$, suggesting the formation of percolated networks among
COMs of ring chains.
%Furthermore, 
%the density dependence of 
%$\lambda(\alpha)$ becomes less pronounced as $\varepsilon_\theta$ increases.
Furthermore, the length scale of $\alpha$ exhibiting a plateau of $\lambda \approx 1$ 
approximately corresponds to 
the characteristic core length, analogous to the
behavior of ring polymers modeled as soft macromolecules.
As $\varepsilon_\theta$ increases, this core length is reduced, 
as shown in Fig.~\ref{fig:ph_com}.
These findings indicate that in flexible ring chains, the cores are large and
overlap each other, but their COMs are not connected with one
another.
In contrast, for stiff ring chains, 
the smaller cores and relatively larger radius of gyration lead to an
increase in 
the number of virtual bonds between COMs.
Thus, the density and chain stiffness strongly
influence 
the structural and dynamic behavior of ring polymer systems.
See Supporting Information for further discussion on the dependence on
chain stiffness and density dependence of the mean square gyration
of radius $\mean{R_\mathrm{g}^2}$, radial distribution function of COMs, $g(r)$,
and the number of virtual bonds $Z_\mathrm{b}$
defined by the radial distribution function of COMs, 
as shown in Fig.~S1-S3 in Supporting Information, respectively.

The next analysis focuses on the one-dimensional hole, \textit{i.e.} ``loop'' structure,
characterized by PH. 
Specifically, PH is performed on each individual ring
polymer $i$ by using the monomer coordinates as input, which generates
a persistent diagram (PD) denoted as $\mathrm{PD}(i)$.
This analysis reveals the birth and death of topological features such
as loops within the structure of the polymer.
Furthermore, 
the ``life'' of the loop is defined as the vertical distance from the
diagonal line in the PD, denoted as $l = d - b$, which quantifies how
long during the increase of $\alpha$
the loop persists before disappearing.
Thus, larger values of $l$
indicates longer-lived loops, reflecting more stable topological
features of the system against the change of threshold.
Next, we performed PH for $(i, j)$ pairs of ring polymer chains using the set of their
coordinates as input, generating PD denoted as $\mathrm{PD}(i \cup j)$.  
Since threading occurs when the loop of one ring polymer disappears due
to penetration by another polymer, 
$\mathrm{PD}(j \to i) = \mathrm{PD}(i) \backslash  \mathrm{PD}(i \cup
j)$ allows 
us to quantify the loops being threaded~\cite{landuzzi2020Persistence}.
Here, 
the set difference operator 
$\backslash$ represents the subtraction of topological features that
vanish when polymer $j$ interacts with polymer $i$.
In this context, polymer 
$i$ is considered ``passive'' while polymer $j$ is the ``active''
participant in the threading process.
Thus, this approach quantifies the threading structures between pairs of ring polymers.
The probability density distributions of
$\mathrm{PD}(i)$, $\mathrm{PD}(i \cup j)$, and $\mathrm{PD}(j \to i)$
with $\varepsilon_\theta = 1.5$ and $5$ at densities $\rho = 0.1$ and
$0.5$ are shown in Fig.~S4-S7 in Supporting Information.

%%%%%%%%% Fig. 3 %%%%%%%%%%%%%%%%%%%%%%%%%%%%%%%%%%%%%%%%%%%%%%%%%%%%%%%%%%%%%%%%%%%%%%%%%
\begin{figure}[t]
\centering
\includegraphics[width=\textwidth]{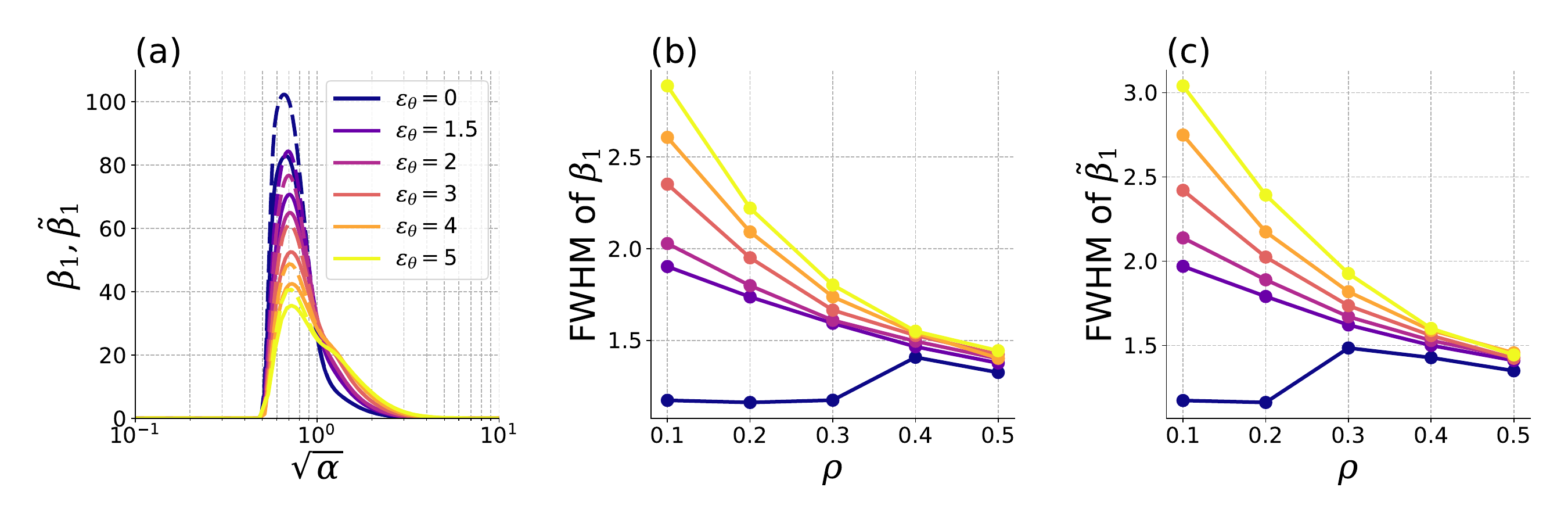}
\caption{
(a) Chain stiffness $\varepsilon_\theta$ dependence of $\beta_1(\alpha)$ and $\tilde{\beta}_1(\alpha)$ at the highest density $\rho=0.5$.
        (b), (c) Plots of full width at half maximum (FWHM) of 
        $\beta_1(\alpha)$ and $\tilde{\beta}_1(\alpha)$ as a function of density $\rho$, respectively.
}
\label{fig:betti1}
\end{figure}
%%%%%%%%%%%%%%%%%%%%%%%%%%%%%%%%%%%%%%%%%%%%%%%%%%%%%%%%%%%%%%%%%%%%%%%%%%%%%%%%%%%%%%%%%%

{\noindent \sffamily \bfseries Chain Stiffness Effects on Loops and Threading in Ring Polymers\\}
To analyze the threading structure by varying the density $\rho$ and
chain stiffness $\varepsilon_\theta$,
we quantify the first Betti number, $\beta_1^{(i)}(\alpha)$, in the $\mathrm{PD}(i)$.
This is defined by 
\begin{equation}
\beta_1^{(i)}(\alpha) = \int_{\alpha}^{\infty} \dd d \int_0^\infty \dd b
 \sum_k \delta (b - b_k^{(i)}) \delta (d - d_k^{(i)}),
\label{eq:1st_Betti}
\end{equation}
where $k$ refers to the $k$-th loop on the 
ring chain $i$.
This $\beta_1(\alpha)$ quantifies the number of loops in the region
where $b< \alpha$ and $d > \alpha$, 
quantifying the number of loops observed at a given $\alpha$.
The average of $\beta_1^{(i)}(\alpha)$ over all ring chains can be
expressed as 
\begin{equation}
\beta_1(\alpha) = \mean{\frac{1}{N} \sum_i \beta_1^{(i)}(\alpha)}.
\end{equation}
The same calculation can be performed for $\mathrm{PD}(j \to i)$, and the
average over all pairs of ring chains $(i, j)$ are denoted as $\tilde\beta_1(\alpha)$.
This $\tilde\beta_1(\alpha)$ measures the number of loops that are being threaded
by other ring chains.
Consequently, it is assured that $\beta_1(\alpha) \ge
\tilde\beta_1(\alpha)$.
Furthermore, $\beta_1(\alpha)$ and $\tilde\beta_1(\alpha)$ converges
asymptotically to
zero with respect to each other as $\alpha$ becomes sufficiently large.

Figure~\ref{fig:betti1}(a) shows the density $\rho$ dependence of 
$\beta_1(\alpha)$ and $\tilde\beta_1(\alpha)$ at the highest density $\rho=0.5$.
The stiff ring exhibits a broader peak at larger length scales
$\alpha$ compared to 
that of the flexible ring, indicating the presence of large loops.
Peaks emerge and decay over similar length scales for both $\beta_1(\alpha)$ and $\tilde\beta_1(\alpha)$.
The peak intensity, however, depends on the chain stiffness $\varepsilon_\theta$.
For sufficiently large $\alpha$, 
the convergence of $\beta_1 \sim \tilde\beta_1$ indicates
that all large loops are involved in threading.
The all results of $\beta_1(\alpha)$ and $\tilde\beta_1(\alpha)$ with varying $\varepsilon_\theta$ and $\rho$
are shown in Fig.~S8 in Supporting Information.

The full width at half maximum (FWHM) of $\beta_1(\alpha)$ and $\tilde\beta_1(\alpha)$ 
are plotted in Fig.~\ref{fig:betti1}(b) and (c).
Note that the FWHM provides insights into average behavior of loop sizes.
The FWHM of $\beta_1(\alpha)$ and $\tilde\beta_1(\alpha)$ are larger for stiff rings 
compared to those of flexible rings, indicating that stiff rings have a broader 
distribution of loop sizes at low density.
As the density increases, the FWHM of $\beta_1(\alpha)$ and $\tilde\beta_1(\alpha)$
converge to a common value.
This result suggests that the stiffness of ring chains significantly influences 
the formation of large loops in dilute solutions,
while the rings strongly overlap and srink the size of loops as the density increases.
Note that 
flexible ring chains with $\varepsilon_\theta = 0$ exhibit non-monotnic behavior at low density $\rho \le 0.3$.
This behavior is attributed to the peak of $\beta_1(\alpha)$ at $\alpha \approx 0.6$, as shown in Fig.~S8(a), 
which is resulted from short-lived loops along the diagonal line in the PD.

%%%%%%%%% Fig. 4 %%%%%%%%%%%%%%%%%%%%%%%%%%%%%%%%%%%%%%%%%%%%%%%%%%%%%%%%%%%%%%%%%%%%%%%%%
\begin{figure}[t]
\centering
\includegraphics[width=\textwidth]{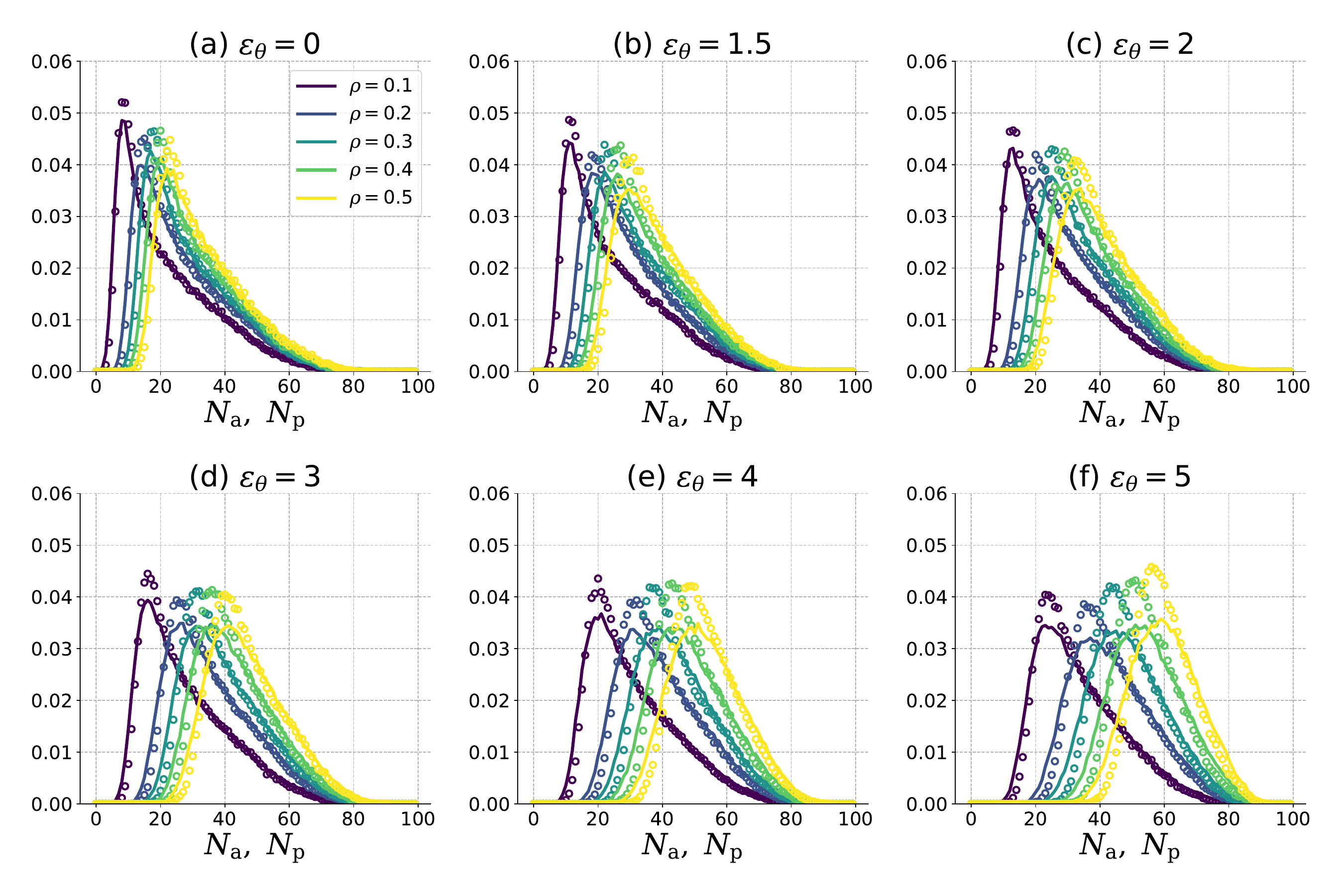}
\caption{Density $\rho$ dependence of probability density distribution of
 active threading number $N_\mathrm{a}$ (points) and
 passive threading number $N_\mathrm{p}$ (solid curves)
by varying the bending energy
 $\varepsilon_\theta = 0$ (a),  $\varepsilon_\theta = 1.5$ (b),
 $\varepsilon_\theta = 2$ (c),  $\varepsilon_\theta = 3$ (d), 
 $\varepsilon_\theta = 4$ (e),  and $\varepsilon_\theta = 5$ (f).
The threshold value is fixed at $l_\mathrm{th}=0$.
}
  \label{fig:na-np-noth}
\end{figure} 
%%%%%%%%%%%%%%%%%%%%%%%%%%%%%%%%%%%%%%%%%%%%%%%%%%%%%%%%%%%%%%%%%%%%%%%%%%%%%%%%%%%%%%%%%%

%%%%%%%%% Fig. 5 %%%%%%%%%%%%%%%%%%%%%%%%%%%%%%%%%%%%%%%%%%%%%%%%%%%%%%%%%%%%%%%%%%%%%%%%%
\begin{figure}[t]
\centering
\includegraphics[width=\textwidth]{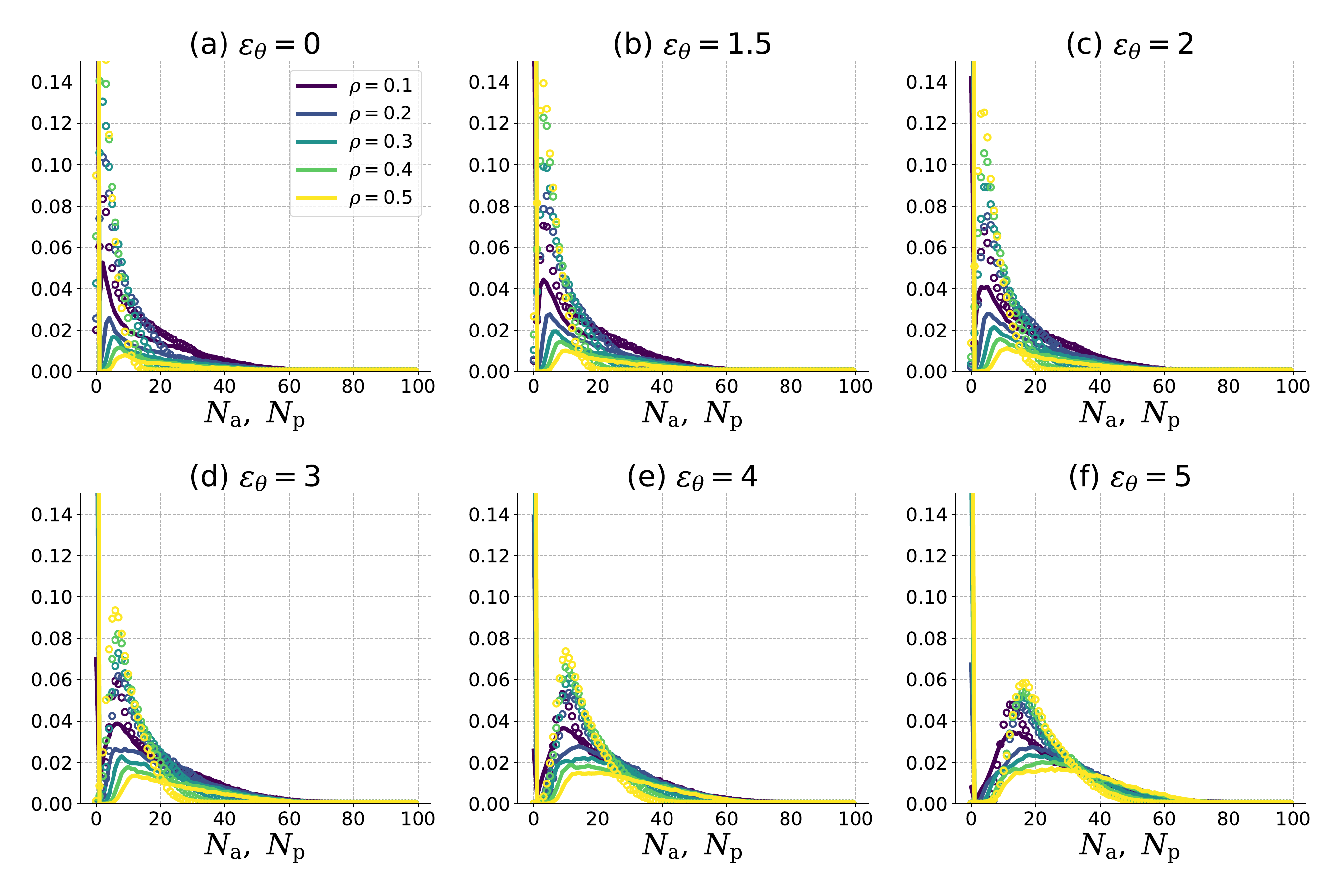}
\caption{Density $\rho$ dependence of 
probability density distribution of
 active threading number $N_\mathrm{a}$ (points) and
 passive threading number $N_\mathrm{p}$ (solid curves)
by varying the bending energy
 $\varepsilon_\theta = 0$ (a),  $\varepsilon_\theta = 1.5$ (b),
 $\varepsilon_\theta = 2$ (c),  $\varepsilon_\theta = 3$ (d), 
 $\varepsilon_\theta = 4$ (e),  and $\varepsilon_\theta = 5$ (f).
The threshold value is fixed at $l_\mathrm{th}=9$.
}
\label{fig:na-np-th}
\end{figure}
%%%%%%%%%%%%%%%%%%%%%%%%%%%%%%%%%%%%%%%%%%%%%%%%%%%%%%%%%%%%%%%%%%%%%%%%%%%%%%%%%%%%%%%%%%

{\noindent \sffamily \bfseries Assymmetry between Active and Passive Threading\\}
We further examine the active threading number 
$N_\mathrm{a}$, 
which represents the number of rings penetrated by a given
ring, and the passive threading number 
$N_\mathrm{p}$, 
which denotes the number of rings that experience penetration by that
same ring.
For the pair $(i, j)$, we define 
\begin{align}
I^{(k)}_{j\rightarrow i} = 
\begin{cases} 
1 & \text{if } l_k \ge l_\mathrm{th} \\
0 & \text{if } l_k < l_\mathrm{th}
\end{cases}
\end{align}
where $l_k$ represents the life of the $k$-th loop in PD$(j \to i)$,
and $l_\mathrm{th}$ is a threshold value for the life used to 
characterize the length scale of threading.
By summing over loop $k$ and polymer $j$ ($i$), the active (passive) threading number,
$N_{\mathrm{a}, j}$ ($N_{\mathrm{p}, j}$) for polymer $j$ ($i$) are obtained, 
expressed as follows:
\begin{align}
N_{\mathrm{a},j} = \sum_i \sum_k I^{(k)}_{j\rightarrow i}, \quad
 N_{\mathrm{p},i} = \sum_j \sum_k I^{(k)}_{j \rightarrow i}.
\end{align}
Furthermore, the averages over all ring chains are denoted by
$N_\mathrm{a}$ and $N_\mathrm{p}$, respectively.
Their statistical averages over all chains ensure
$\mean{N_\mathrm{a}} = \mean{N_\mathrm{p}}$ because, when threading
occurs, active and passive threading are always counted once,
respectively.

Figure~\ref{fig:na-np-noth} presents the
probability density distribution of 
$N_\mathrm{a}$ and $N_\mathrm{p}$, respectively.
Note that $N_\mathrm{a}$ and $N_\mathrm{p}$ were calculated by 
including threading at all length scales, with the threshold
$l_\mathrm{th}$ set to zero.
It is demonstrated that for both $N_\mathrm{a}$ and $N_\mathrm{p}$, 
the peak shifts to higher values with increasing chain stiffness 
$\varepsilon_\theta$ and density 
$\rho$, indicating a greater occurrence of threading.
Notably, the density dependence of the distribution becomes more pronounced
for stiff rings compared to that of flexible rings.
In addition, $N_\mathrm{p}$ exhibits a slightly broader
distribution 
than $N_\mathrm{a}$ at high density for stiff rings.
This asymmetric property between $N_\mathrm{a}$ and $N_\mathrm{p}$ was
found to be pronounced for longer stiff rings, 
suggesting that the 
passive threading is significantly influenced by 
the presence of long-lived loops.
In other words, larger loops are likely to be involved in the passive threading.

We further characterize the long-lived active and passive threading structures by
introducing the threshold value $l_\mathrm{th}$, which has a dimension $\sigma^2$.
Since points near the diagonal line are considered noisy, we introduce
$l_\mathrm{th}$ to filter out threading associated with loops of short
life, thereby characterizing loops that are mostly correlated with
topological constraints.
While the results for varying $l_\mathrm{th}$ are not displayed, the
threshold value $l_\mathrm{th} = 9$ was determined to capture the most
relevant characteristics, and the corresponding results are shown below.

Figure~\ref{fig:na-np-th} illustrates the density
dependence of probability density
distribution of active and passive threading numbers, $N_\mathrm{a}$ and
$N_\mathrm{b}$, at $l_\mathrm{th}=9$.
For flexible ring chains, both $N_\mathrm{a}$ and $N_\mathrm{p}$ show
the tendency of the decrease toward
zero as the density $\rho$ increases.
This trend is expected to become more pronounced as the threshold
value $l_\mathrm{th}$ increases.
This observation suggests that the number of loops necessary for
threading becomes minimal in higher densities, consistent with the
overlapping structures between
the crumbled globules characteristic of flexible ring chains.
In contrast, for stiff ring chains, the distribution of $N_\mathrm{a}$
exhibit a peak at $N_\mathrm{a} \approx 20$ across all densities, whereas
the distribution of $N_\mathrm{p}$ shows two distinct peaks, one at
$N_\mathrm{p}=0$ and another at 
$N_\mathrm{p} \approx 20$.
In addition, the latter peak broadens as the density $\rho$ increases.
This observation implies that, when focusing on passive threading of
stiff ring chains, they can 
be categorized into two different types: those having
large loops facilitate threading and those lacking such structures.
The latter rings are regarded as exhibiting more compact characteristic
rather than those of the former.

{\noindent \sffamily \bfseries Conclusion\\}
%%%%%%%% Conclusions %%%%%%%%
In summary, we employed PH analysis to characterize threading from 
MD simulations of the KG model for ring polymers.
Specifically, we focused on the threading structure as influenced by the
density $\rho$ and chain stiffness $\varepsilon_\theta$, while
maintaining the chain length of $N=400$.
Our analyses consists of three components:
First, we examined the zero-th Betti number $\beta_0(\alpha)$ to
quantify the number of connected components formed by COMs of the polymers.
This analysis demonstrates that numerous small clusters of COMs persist for flexible ring
chains even at high densities, whereas a percolated network of COMs
develops for stiff ring chains as the density increases.
Second, we calculated the first Betti numbers, $\beta_1(\alpha)$ and
$\tilde\beta_1(\alpha)$, from $\mathrm{PD}(j \to i)$ to
characterize the threading structure between pairs of ring chains.
It is shown that stiff ring chains exhibit large-scale loops that
facilitate
threading as the density $\rho$ increases.
Furthermore, we also computed the active and passive threading numbers, $N_\mathrm{a}$
and $N_\mathrm{p}$.
As both $\varepsilon_\theta$ and $\rho$ increase, their averages
become larger, indicating greater generations of threading, accompanied by
the asymmetric behavior of the distributions of $N_\mathrm{a}$
and $N_\mathrm{p}$.
Finally, 
we introduced the threshold value $l_\mathrm{th}$ to emphasize 
long-lived threading structures in the calculations of $N_\mathrm{a}$
and $N_\mathrm{p}$.
This analysis reveals that the distributions of $N_\mathrm{a}$ and
$N_\mathrm{p}$ converges to zero for flexible ring chains as the density
increases.
In contract, for stiff ring chains, the distribution of $N_\mathrm{p}$
bifurcates into two
distinct peaks, indicating heterogeneous threading structure
characterized by 
rings with large-scale loops that facilitate threading and those that
have compact ring characteristic.
This heterogeneous threading structure observed in stiff ring chains
serves as the underlying mechanism for topological glasses, which exhibit
heterogeneous rearrangement dynamics of COMs analogous to those
of glass-forming liquids.

\begin{acknowledgement}
This work was supported by 
JSPS KAKENHI Grant-in-Aid 
Grant Nos.~\mbox{JP24H01719}, \mbox{JP22H04542}, \mbox{JP22K03550},
 \mbox{JP23K27313}, and \mbox{JP23H02622}
We also acknowledge
the Fugaku Supercomputing Project (Nos.~JPMXP1020230325 and JPMXP1020230327) and 
the Data-Driven Material Research Project (No.~\mbox{JPMXP1122714694})
from the
Ministry of Education, Culture, Sports, Science, and Technology and to
 Maruho Collaborative Project for Theoretical Pharmaceutics.
The numerical calculations were performed at Research Center for
Computational Science, Okazaki Research Facilities, National Institutes
 of Natural Sciences (Project: \mbox{24-IMS-C051}).
DM thanks the Royal Society for support through a University Research Fellowship and 
 the European Research Council (ERC) under the European Union's Horizon 2020 research and innovation program (grant agreement No 947918, TAP).
\end{acknowledgement}

%\begin{suppinfo}
%
%A listing of the contents of each file supplied as Supporting Information
%should be included. For instructions on what should be included in the
%Supporting Information as well as how to prepare this material for
%publications, refer to the journal's Instructions for Authors.
%
%The following files are available free of charge.
%\begin{itemize}
%  \item Filename: brief description
%  \item Filename: brief description
%\end{itemize}
%
%\end{suppinfo}

%\bibliography{ring}
\providecommand{\latin}[1]{#1}
\makeatletter
\providecommand{\doi}
  {\begingroup\let\do\@makeother\dospecials
  \catcode`\{=1 \catcode`\}=2 \doi@aux}
\providecommand{\doi@aux}[1]{\endgroup\texttt{#1}}
\makeatother
\providecommand*\mcitethebibliography{\thebibliography}
\csname @ifundefined\endcsname{endmcitethebibliography}
  {\let\endmcitethebibliography\endthebibliography}{}

\end{document}

%\begin{tocentry}
%
%
%\end{tocentry}

%\begin{abstract}
%\end{abstract}

\section{Radius of gyration}

For flexible ring polymers with the chain stiffness
$\varepsilon_\theta=0$ in semidilute solutions, molecular dynamics (MD)
simulations using the Kremer--Grest (KG) model reveal that the mean square radius of gyration
$\mean{R_\mathrm{g}^2}$ as a function of 
density $\rho$ follows the
scaling behavior of $\mean{R_\mathrm{g}^2} \sim \rho^{-0.59}$~\cite{cai2022Conformation}.
More specifically, using
the mean square radius of gyration in the dilute limit, denoted as
$\mean{R_\mathrm{g}^{\circ 2}}$, and 
the overlap density $3N / (4\pi \langle R_\mathrm{g}^{\circ
2}\rangle^{3/2})$, the scaling relation
\begin{equation}
\frac{\langle R_\mathrm{g}^2 \rangle}{\langle R_\mathrm{g}^{\circ 2}
 \rangle} = \left[1 + a\left(\frac{\rho} { \rho^*} \right)\right]^{b}, 
\label{eq:rg-mastercurve2}
\end{equation}
was proposed.
Here, $a = 0.45$ and $b = -0.59$ were the fitting parameters.

\begin{figure}[H]
\centering
\includegraphics[width=0.6\textwidth]{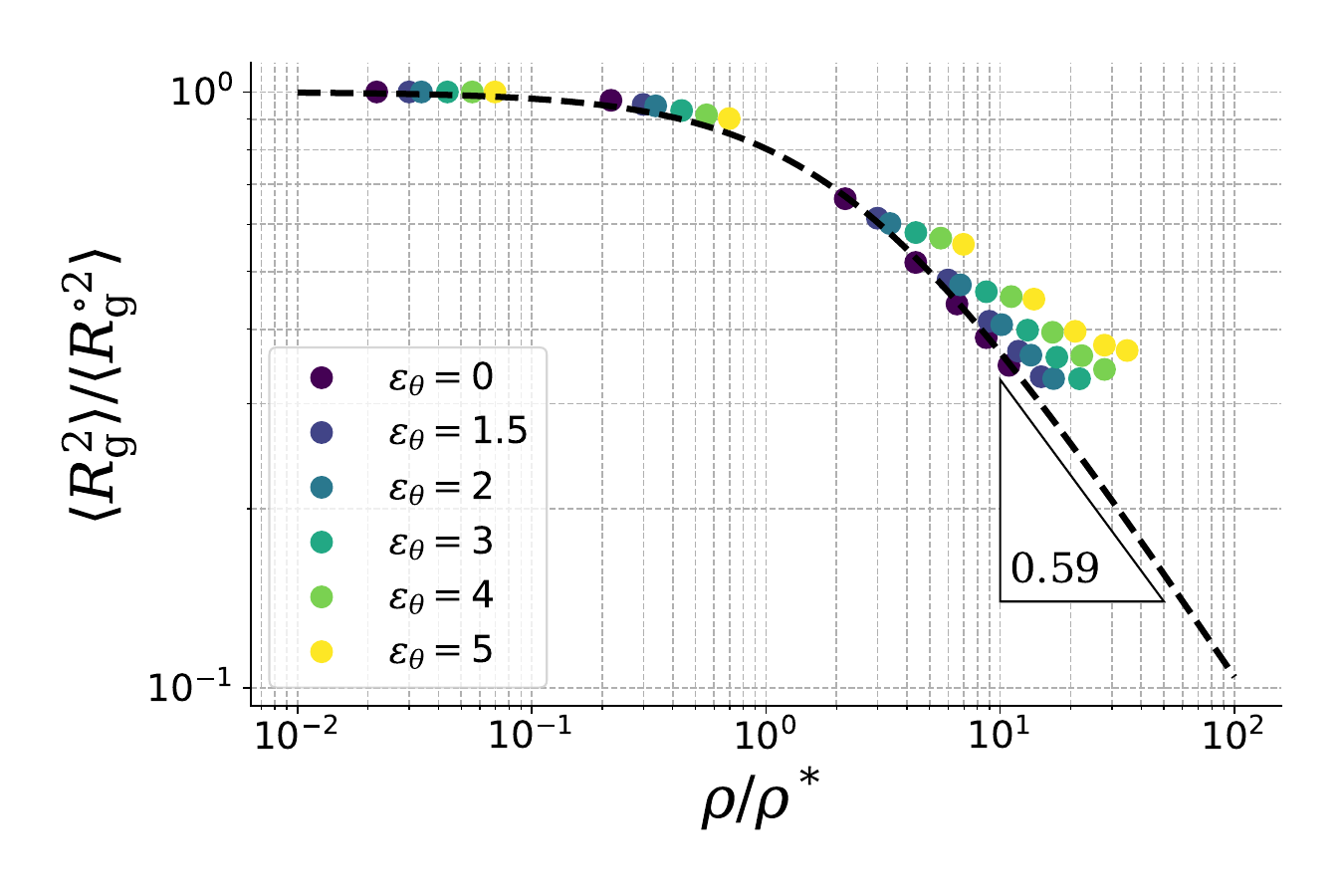}
\caption{density $\rho$ scaled by the overlap density $\rho^*$ dependence of
mean square radius of gyration $\mean{R_\mathrm{g}^2}$ scaled by its
 dilute limit value $\mean{R_\mathrm{g}^{\circ 2}}$.
}
\label{fig:rg}
\end{figure}

Figure~\ref{fig:rg} shows the chain stiffness $\varepsilon_\theta$
dependence of the relationship between 
$\langle R_\mathrm{g}^2 \rangle/\langle R_\mathrm{g}^{\circ 2}\rangle$
and $\rho/ \rho^*$ from our MD simulations.
Note that 
$\mean{R_\mathrm{g}^{\circ 2}}$ was calculated at $\rho = 0.001$.
The results reveal that $\mean{R_\mathrm{g}^2}$ decreases with
increasing density $\rho$ and exhibits significant deviation from the
scaling of Eq.~\eqref{eq:rg-mastercurve2} 
when chain stiffness $\varepsilon_\theta$ is large, 
particularly noticeable for $\rho/\rho^* >1$.This deviation from Eq.~\eqref{eq:rg-mastercurve2} implies that the
influence of density $\rho$ on
chain conformation varies depending on the chain stiffness $\varepsilon_\theta$.

%%%%%%%%% Fig. S1 %%%%%%%%%%%%%%%%%%%%%%%%%%%%%%%%%%%%%%%%%%%%%%%%%%%%%%%%%%%%%%%%%%%%%%%%%
%%%%%%%%%%%%%%%%%%%%%%%%%%%%%%%%%%%%%%%%%%%%%%%%%%%%%%%%%%%%%%%%%%%%%%%%%%%%%%%%%%%%%%%%%%

\section{Radial distribution function of cneter of mass}

\begin{figure}[H]
  \centering
  \includegraphics[width=0.9\textwidth]{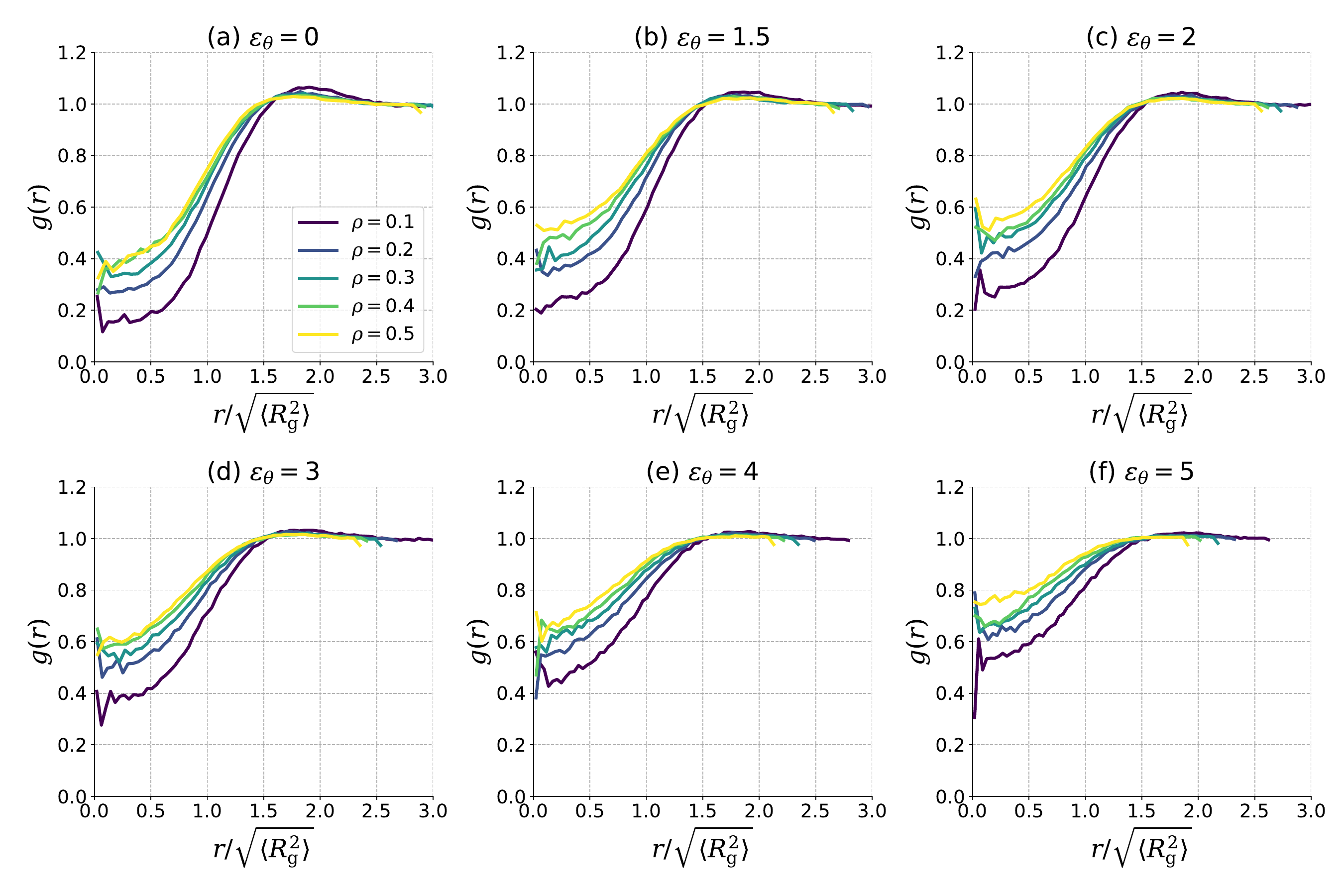}
  \caption{
Radial distribution function $g(r)$ for COM of ring polymers as a
 function of the scaled distance $r/\sqrt{\mean{R_\mathrm{g}^2}}$ at
 $\varepsilon_\theta = 0$ (a), $\varepsilon_\theta =1.5$ (b),
 $\varepsilon_\theta = 2$ (c), $\varepsilon_\theta = 3$ (d),
 $\varepsilon_\theta = 4$ (e), and $\varepsilon_\theta = 5$ (f).
}
\label{fig:rdf}
\end{figure}

We calculated the radial distribution function, $g(r)$, for center of
mass (COM) of ring chains.
The results are illustrated in Fig.~\ref{fig:rdf}.
As demonstrated in Fig.~\ref{fig:rdf}, $g(r)$ exhibits finite values at the length scale
$r<\sqrt{\mean{R_\mathrm{g}^2}}$, 
indicating significant interpenetration between the ring chains. 
For flexible ring polymers with $\varepsilon_\theta = 0$, 
$g(r)$ broadens with increasing density $\rho$, suggesting that
the chains become less spatially separated from one another.
In addition, for stiff ring polymers with $\varepsilon_\theta = 5$, 
the degree of interpenetration becomes more pronounced as the density
$\rho$ increases.
This observation is attributed to
the larger mean square radius of gyration, $\mean{R_\mathrm{g}^2}$,
compared to that of flexible ring chains with $\varepsilon_\theta = 0$
at the same $\rho$ for dense systems (see Fig.~\ref{fig:rg}).
An analogous observation with respect to the chain stiffness and density dependence of $g(r)$ for ring polymers
was reported in a previous study~\cite{bernabei2013Fluids}.

%%%%%%%%% Fig. S2 %%%%%%%%%%%%%%%%%%%%%%%%%%%%%%%%%%%%%%%%%%%%%%%%%%%%%%%%%%%%%%%%%%%%%%%%%
%%%%%%%%%%%%%%%%%%%%%%%%%%%%%%%%%%%%%%%%%%%%%%%%%%%%%%%%%%%%%%%%%%%%%%%%%%%%%%%%%%%%%%%%%%

\section{Virtual bond number}

As shown in Figs.~\ref{fig:rg} and \ref{fig:rdf}, the influence of
density $\rho$ 
on $\mean{R_\mathrm{g}^2}$ and $g(r)$ significantly varies with chain
stiffness $\varepsilon_\theta$.
To characterized the connectivity between COMs by varying $\rho$ and
$\varepsilon_\theta$, we introduced a virtual bond between ring polymers
$i$ and $j$.
Specifically, if the distance between the COMs of polymers $i$ and $j$, denoted as
$r_{ij}$, satisfies
\begin{equation}
  r_{ij} \le \sqrt{\langle R_\mathrm{g}^2 \rangle},
  \label{eq:vb}
\end{equation}
the two polymer chains are considered to be virtually bonded~\cite{goto2023Unraveling}.
The number of virtual bond is denoted as $Z_\mathrm{b}$.
The average number of virtual bonds can be expressed by 
\begin{align}
\langle Z_\mathrm{b}\rangle = \int_0^{\sqrt{\langle R_\mathrm{g}^2
 \rangle}} 4\pi r^2 \left(\frac{\rho}{N}\right) g(r) \dd r.
 \label{eq:vb-rdf}
\end{align}
Note that the threshold of the virtual bond is less than 
the contact distance, $2\sqrt{\mean{R_\mathrm{g}^2}}$, to emphasize the
overlapping between COMs.

\begin{figure}[H]
  \centering
  \includegraphics[width=0.6\textwidth]{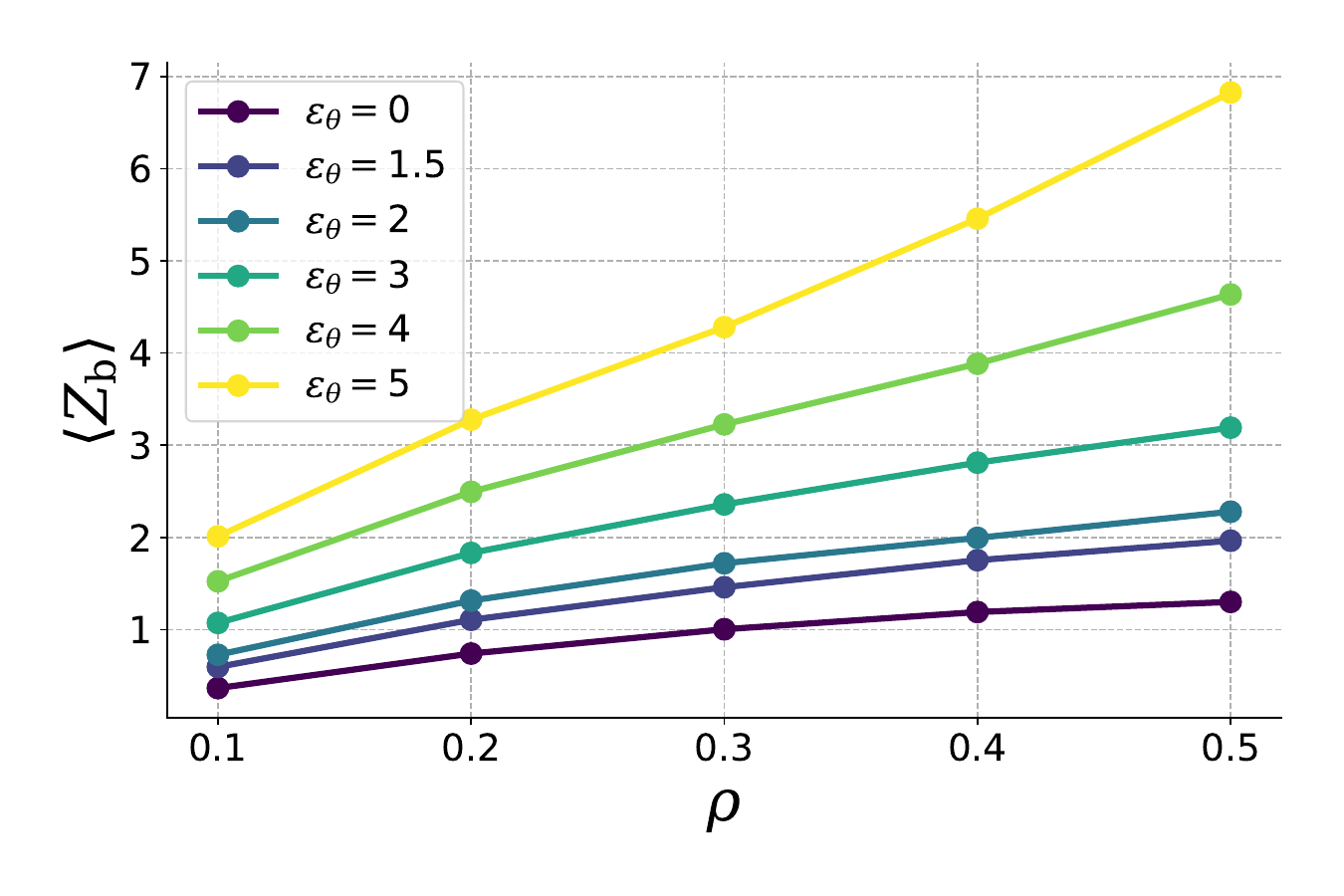}
  \caption{
Density $\rho$ dependence of the average number of virtual bonds,
 $\mean{Z_\mathrm{b}}$ by varying the chain stiffness $\varepsilon_\theta$.
}
\label{fig:Zb}
\end{figure}

Figure~\ref{fig:Zb} shows $\langle Z_\mathrm{b} \rangle$ as a function
of density $\rho$ by varying the chain stiffness $\varepsilon_\theta$.
The $\mean{Z_\mathrm{b}}$ is an increasing function of $\rho$.
As $\varepsilon_\theta$ increases, the slope becomes steeper,
indicating a greater dependence on $\rho$.
In contrast, for flexible ring chains with $\varepsilon_\theta = 0$, 
$g(r/\sqrt{\mean{R_\mathrm{g}^2}})$ was found to saturate with
increasing $\rho$, as demonstrated in the previous
study~\cite{cai2022Conformation}. 
Similarly, 
$\mean{Z_\mathrm{b}}$ is also expected to approach saturation towards
a finite value.
This distinction in the density $\rho$ dependence on the average number of
virtual bonds $\mean{Z_\mathrm{b}}$ implies a 
significant difference in 
intermolecular
interaction between ring polymers as the chain stiffness
$\varepsilon_\theta$ varies.

%%%%%%%%% Fig. S3 %%%%%%%%%%%%%%%%%%%%%%%%%%%%%%%%%%%%%%%%%%%%%%%%%%%%%%%%%%%%%%%%%%%%%%%%%
%%%%%%%%%%%%%%%%%%%%%%%%%%%%%%%%%%%%%%%%%%%%%%%%%%%%%%%%%%%%%%%%%%%%%%%%%%%%%%%%%%%%%%%%%%

\section{Persistent diagram}

\begin{figure}[H]
\centering
\includegraphics[width=0.9\textwidth]{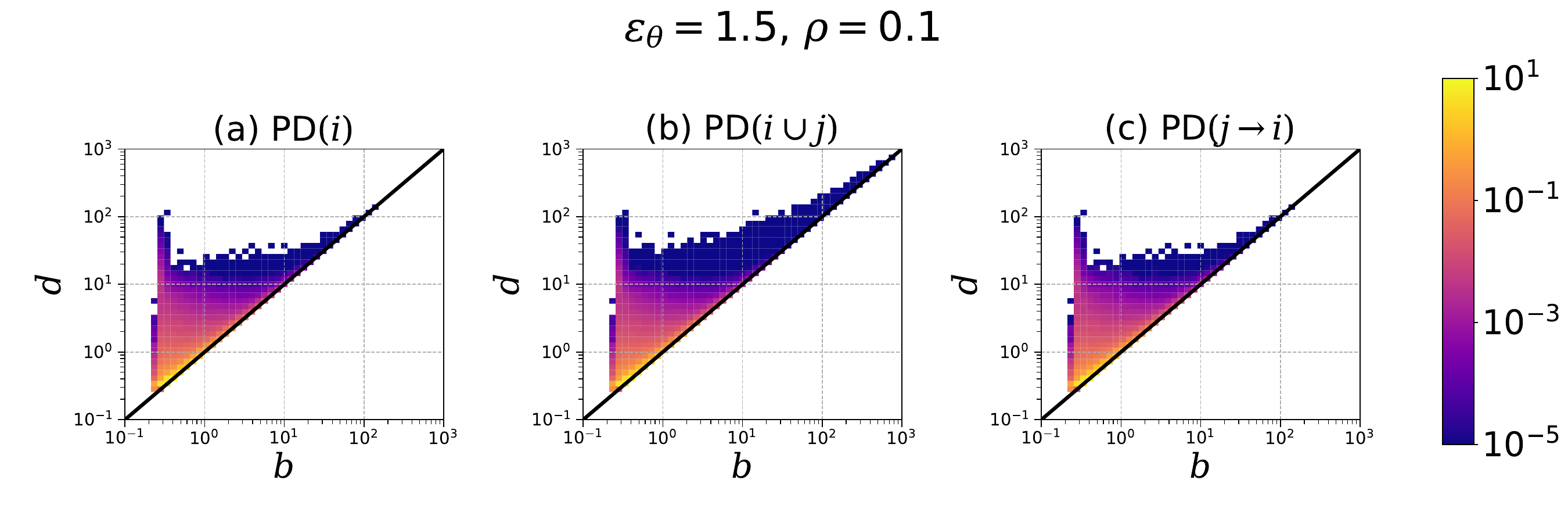}
\caption{Persistent diagrams, $\mathrm{PD}(i)$ (a), $\mathrm{PD}(i\cup
 j)$, and $\mathrm{PD}(i\to j)$, with $\varepsilon = 1.5$ and $\rho=0.1$.}
\label{fig:pd_rho0.1_bend1.5}
\end{figure}

\begin{figure}[H]
\centering
\includegraphics[width=0.9\textwidth]{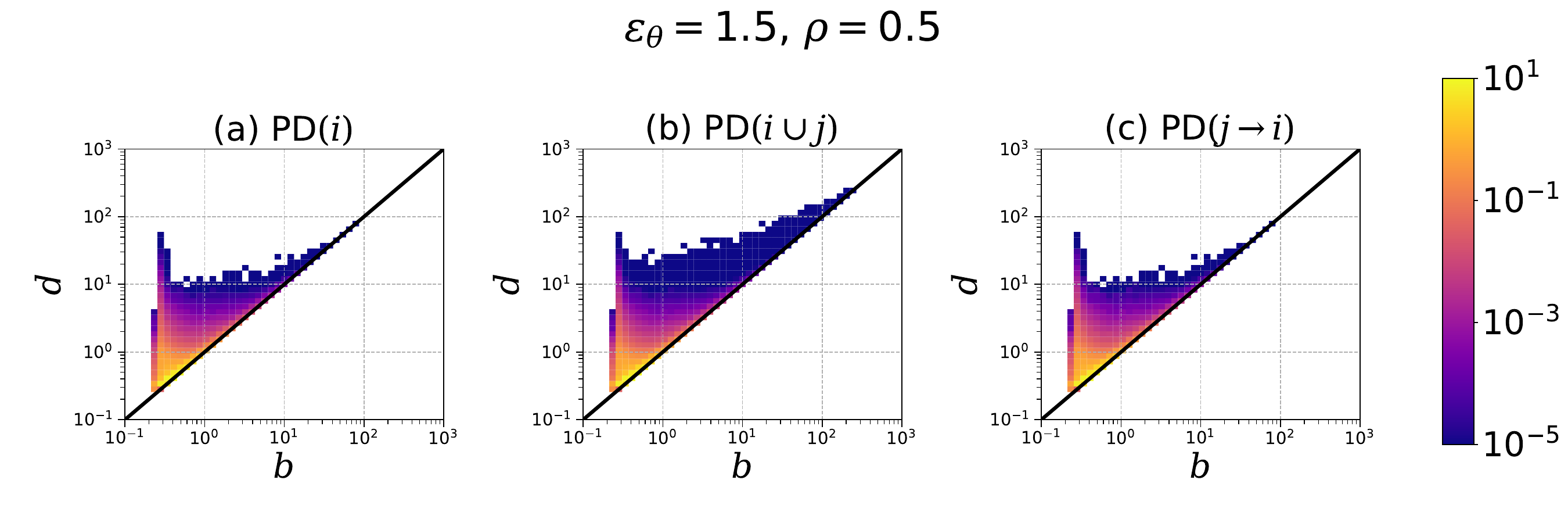}
\caption{Persistent diagrams, $\mathrm{PD}(i)$ (a), $\mathrm{PD}(i\cup
 j)$, and $\mathrm{PD}(i\to j)$, with $\varepsilon = 1.5$ and $\rho=0.5$.}
\label{fig:pd_rho0.5_bend1.5}
\end{figure}

\begin{figure}[H]
\centering
\includegraphics[width=0.9\textwidth]{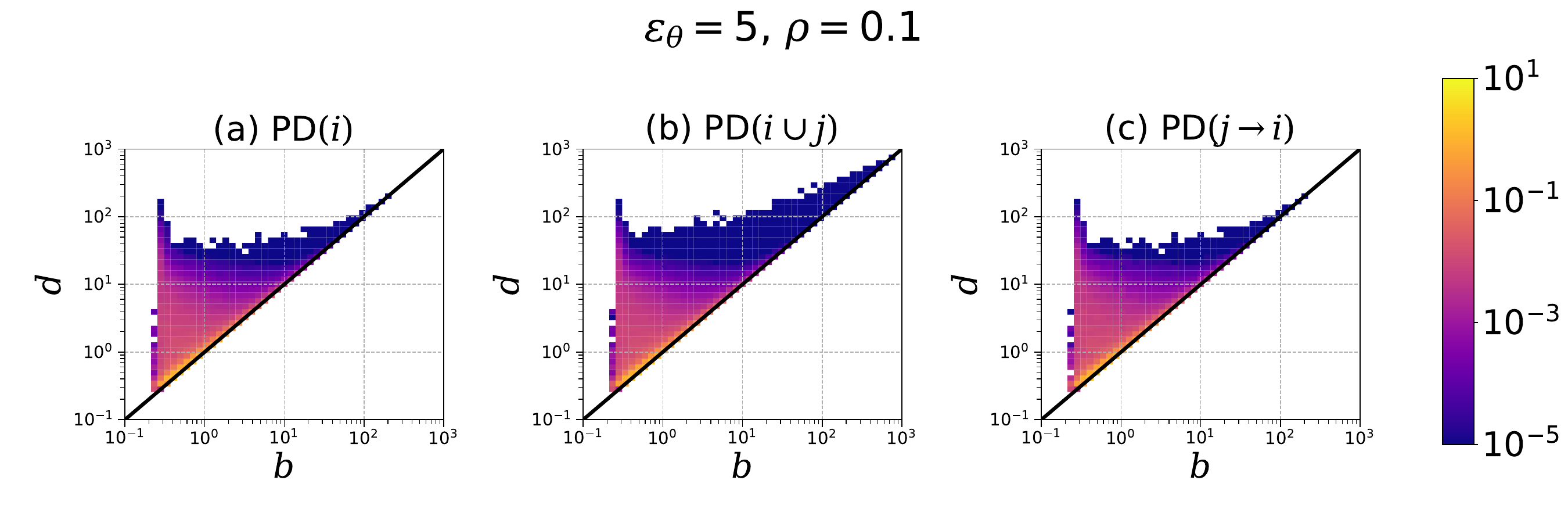}
\caption{Persistent diagrams, $\mathrm{PD}(i)$ (a), $\mathrm{PD}(i\cup
 j)$, and $\mathrm{PD}(i\to j)$, with $\varepsilon = 5$ and $\rho=0.1$.}
\label{fig:pd_rho0.1_bend5}
\end{figure}

\begin{figure}[H]
\centering
\includegraphics[width=0.9\textwidth]{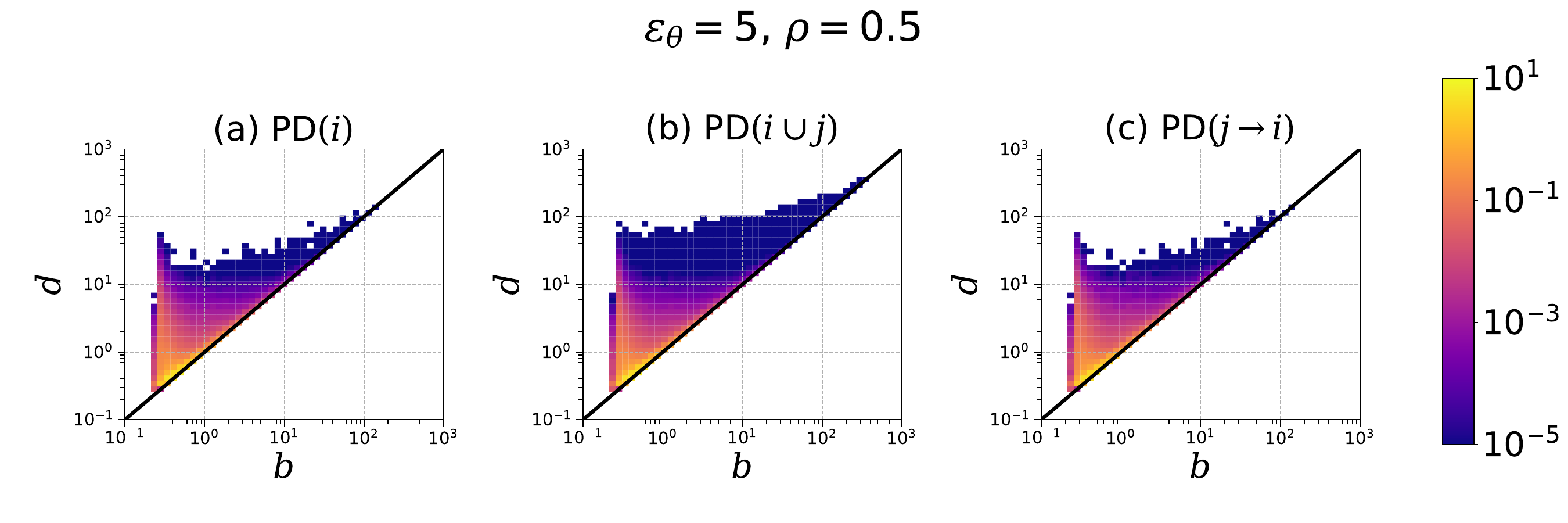}
\caption{Persistent diagrams, $\mathrm{PD}(i)$ (a), $\mathrm{PD}(i\cup
 j)$, and $\mathrm{PD}(i\to j)$, with $\varepsilon = 5$ and $\rho=0.5$.}
\label{fig:pd_rho0.5_bend5}
\end{figure}

Figures~\ref{fig:pd_rho0.1_bend1.5}, \ref{fig:pd_rho0.5_bend1.5},
\ref{fig:pd_rho0.1_bend5}, and \ref{fig:pd_rho0.5_bend5} present the probability density distributions of
$\mathrm{PD}(i)$, $\mathrm{PD}(i \cup j)$, and $\mathrm{PD}(j \to i)$ at
combinations of $(\varepsilon_\theta, \rho) = (1.5, 0.1)$, $(1.5, 0.5)$,
$(5, 0.1)$, and $(5, 0.5)$, respectively.
The general shape of $\mathrm{PD}(i)$ remains consistent
regardless of variations in 
density $\rho$ or chain hardness $\varepsilon_\theta$.
%The main portion of the distribution follows the diagonal line in the birth-death space, 
The area with the highest frequency appears close to the diagonal
with a prominent vertical 
distribution at $b \approx 0.22$.
The distribution along the diagonal line represents loops
that are formed and quickly disappear.
These short-lived loops, characterized by small values of life $l$,
are typically
regarded as noise because they do not
significantly contribute to threading structures.
In contrast, the distribution along $b \approx 0.22$ is interpreted as
loops generated by the inherent stiffness of the polymer chain backbone.
Specifically, this value corresponds to the characteristic loop size related to the
average bond length, $l_\mathrm{b} = 0.965 \approx 2\sqrt{0.22}$ of the
KG model.
The loops observed in the intermediate region, between the diagonal line
and $b\approx 0.22$, are thought to be associated with secondary
structures~\cite{hiraoka2016Hierarchical}, such as the folding or compact configurations of ring
polymers.
These loops arise from internal conformational changes, bringing
different parts of the polymer chain closer together, forming transient or
quasi-stable folded structures.
Unlike short-lived loops near the diagonal line, 
these intermediate loops contribute to the overall topological complexity of the system.

In the 
$\mathrm{PD}(i \cup j)$,
the distribution along the diagonal is more extended compared to that of 
$\mathrm{PD}(i)$. 
In addition, the intermediate distribution exhibits a more elongated
shape. 
This is attributed to the creation of new loops caused by the contact
between pairs of ring chains.
These newly formed loops arise from the threading of
ring chains, leading to an increase in the complexity of the
structures, characterized by longer life $l$, due to the
interaction between different chains.
%Furthermore, the elongation is more pronounced with decreasing the density
%$\rho$ and chain stiffness $\varespsilon_\theta$.
%In contrast, the distribution along 
%$b\approx 0.22$ remains mostly unchanged. 
%However, a slight increase is observed in the region, where the life time 
%$l$ becomes larger.
%This increase is attributed to the influence of ring polymer pairs
%coming into contact, which enhances the persistence of certain loops. 
The distribution 
$\mathrm{PD}(j\to i)$, representing the difference between 
$\mathrm{PD}(i)$ and 
$\mathrm{PD}(i\cup j)$, does not exhibit significant changes compared to the shape of 
$\mathrm{PD}(i)$. 
This indicates that the loops of one ring polymer are significantly
influenced by the threading interaction with other ring polymers.
A detailed discussion regarding the chain stiffness $\varepsilon_\theta$
and density $\rho$ is provided in the main text, 
where the analysis of the zero-th and first Betti numbers offers further insights.

%%%%%%%%% Fig. 4 %%%%%%%%%%%%%%%%%%%%%%%%%%%%%%%%%%%%%%%%%%%%%%%%%%%%%%%%%%%%%%%%%%%%%%%%%
%%%%%%%%%%%%%%%%%%%%%%%%%%%%%%%%%%%%%%%%%%%%%%%%%%%%%%%%%%%%%%%%%%%%%%%%%%%%%%%%%%%%%%%%%%

%%%%%%%%% Fig. 5 %%%%%%%%%%%%%%%%%%%%%%%%%%%%%%%%%%%%%%%%%%%%%%%%%%%%%%%%%%%%%%%%%%%%%%%%%
%%%%%%%%%%%%%%%%%%%%%%%%%%%%%%%%%%%%%%%%%%%%%%%%%%%%%%%%%%%%%%%%%%%%%%%%%%%%%%%%%%%%%%%%%%

%%%%%%%%% Fig. 6 %%%%%%%%%%%%%%%%%%%%%%%%%%%%%%%%%%%%%%%%%%%%%%%%%%%%%%%%%%%%%%%%%%%%%%%%%
%%%%%%%%%%%%%%%%%%%%%%%%%%%%%%%%%%%%%%%%%%%%%%%%%%%%%%%%%%%%%%%%%%%%%%%%%%%%%%%%%%%%%%%%%%

%%%%%%%%% Fig. 7 %%%%%%%%%%%%%%%%%%%%%%%%%%%%%%%%%%%%%%%%%%%%%%%%%%%%%%%%%%%%%%%%%%%%%%%%%
%%%%%%%%%%%%%%%%%%%%%%%%%%%%%%%%%%%%%%%%%%%%%%%%%%%%%%%%%%%%%%%%%%%%%%%%%%%%%%%%%%%%%%%%%%

\section{First Betti number}

\begin{figure}[H]
\centering
\includegraphics[width=0.9\textwidth]{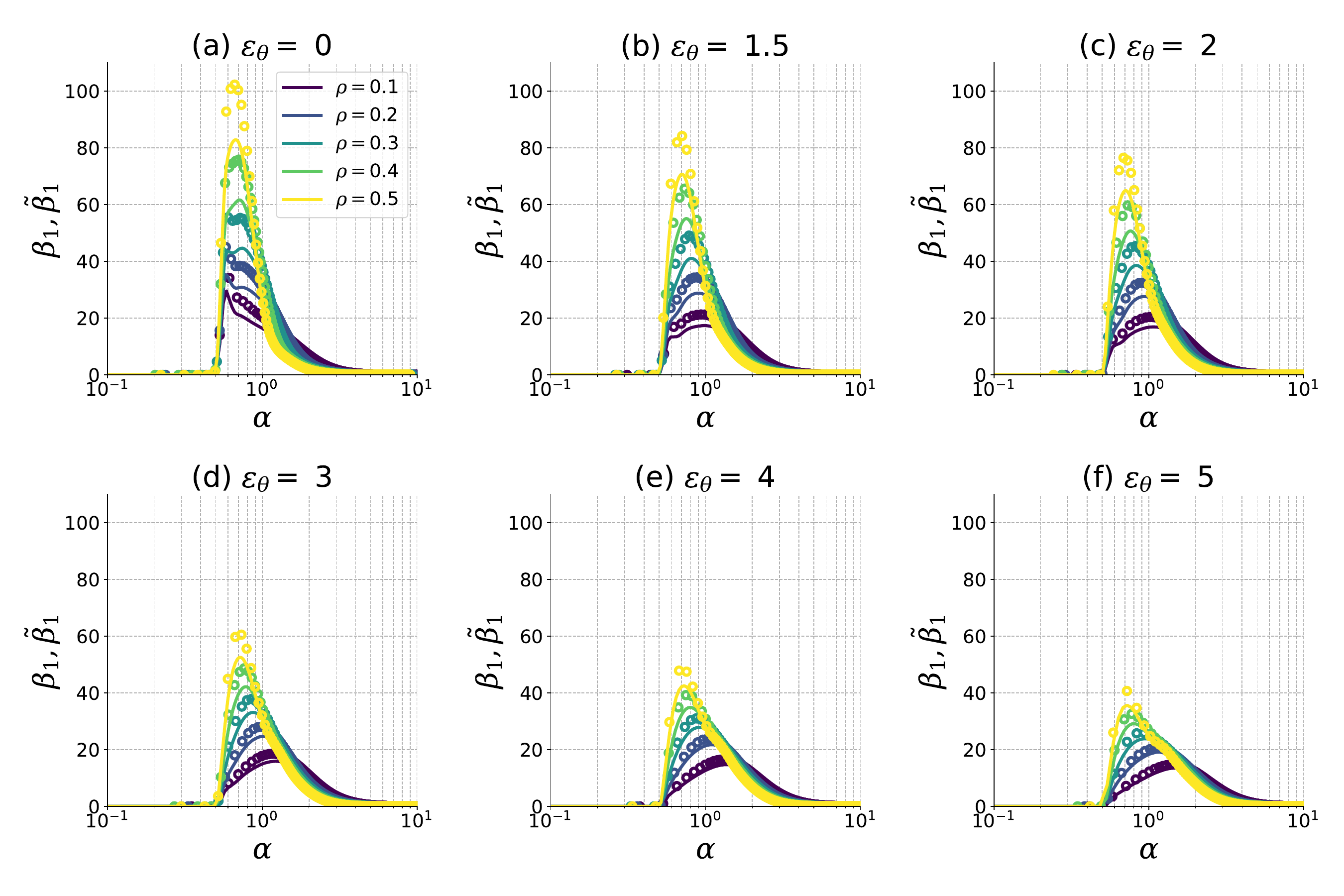}
\caption{Density $\rho$ dependence of $\beta_1(\alpha)$ (points) and $\tilde{\beta}_1(\alpha)$ (solid
 curves)
by varying the bending energy
 $\varepsilon_\theta = 0$ (a),  $\varepsilon_\theta = 1.5$ (b),
 $\varepsilon_\theta = 2$ (c),  $\varepsilon_\theta = 3$ (d), 
 $\varepsilon_\theta = 4$ (e),  and $\varepsilon_\theta = 5$ (f).
}
\label{fig:betti1-si}
\end{figure}

Figure~\ref{fig:betti1-si} shows $\beta_1(\alpha)$ and
$\tilde\beta_1(\alpha)$ for varying 
$\rho$ and $\varepsilon_\theta$.
%The stiff ring exhibits a broader peak at larger length scales
%$\alpha$ compared to 
%that of the flexible ring, indicating the presence of large loops.
As the density increases, this peak sharpens, with its position shifting to 
smaller $\alpha$, signifying the formation of smaller loops.
However, the discrepancy in $\beta_1(\alpha)$ and
$\tilde\beta_1(\alpha)$
between flexible and stiff rings becomes more pronounced with increasing density $\rho$.
This observation aligns with 
the fact that 
the density $\rho$ dependence of
the mean square
radius of gyration $\mean{R_\mathrm{g}^2}$ 
exceeds the expected scaling behavior of $\mean{R_\mathrm{g}^2} \sim \rho^{-0.59}$ for stiff rings
as $\rho$ increases (see Fig.~\ref{fig:rg}).
%%%%%%%%% Fig. 8 %%%%%%%%%%%%%%%%%%%%%%%%%%%%%%%%%%%%%%%%%%%%%%%%%%%%%%%%%%%%%%%%%%%%%%%%%
%%%%%%%%%%%%%%%%%%%%%%%%%%%%%%%%%%%%%%%%%%%%%%%%%%%%%%%%%%%%%%%%%%%%%%%%%%%%%%%%%%%%%%%%%%

%\begin{acknowledgement}
%
%
%
%\end{acknowledgement}

%\begin{suppinfo}
%
%A listing of the contents of each file supplied as Supporting Information
%should be included. For instructions on what should be included in the
%Supporting Information as well as how to prepare this material for
%publications, refer to the journal's Instructions for Authors.
%
%The following files are available free of charge.
%\begin{itemize}
%  \item Filename: brief description
%  \item Filename: brief description
%\end{itemize}
%
%\end{suppinfo}

%\bibliography{ring}
\providecommand{\latin}[1]{#1}
\makeatletter
\providecommand{\doi}
  {\begingroup\let\do\@makeother\dospecials
  \catcode`\{=1 \catcode`\}=2 \doi@aux}
\providecommand{\doi@aux}[1]{\endgroup\texttt{#1}}
\makeatother
\providecommand*\mcitethebibliography{\thebibliography}
\csname @ifundefined\endcsname{endmcitethebibliography}
  {\let\endmcitethebibliography\endthebibliography}{}